\documentclass[prb,twocolumn,showpacs,preprintnumbers,amsmath,amssymb,floatfix]{revtex4}
\usepackage{epsfig}
\usepackage{graphicx}
\usepackage{color}

\begin{document}

\title{Adiabatic quantum dynamics of the Lipkin-Meshkov-Glick model}

\author{Tommaso Caneva}
\affiliation{International School for Advanced Studies (SISSA), Via Beirut 2-4,
  I-34014 Trieste, Italy}

\author{Rosario Fazio}
\affiliation{International School for Advanced Studies (SISSA), Via Beirut 2-4,
  I-34014 Trieste, Italy}
\affiliation{NEST-CNR-INFM $\&$ Scuola Normale Superiore di Pisa, 
  Piazza dei Cavalieri 7, 56126 Pisa, Italy}

\author{Giuseppe E. Santoro}
\affiliation{International School for Advanced Studies (SISSA), 
  Via Beirut 2-4, I-34014 Trieste, Italy}
\affiliation{CNR-INFM Democritos National Simulation Center, 
  Via Beirut 2-4, I-34014 Trieste, Italy}
\affiliation{International Centre for Theoretical Physics (ICTP), 
  P.O.Box 586, I-34014 Trieste, Italy}

\date{\today}

\begin{abstract}
%{
%\color{red}
We study the adiabatic dynamics of the Lipkin-Meshkov-Glick (LMG) close to its quantum critical 
point by linearly switching the transverse field from an initial large value to zero.   
We concentrate our attention on the residual energy after the quench in order to characterize 
the level of diabaticity of the evolution. As a function of the characteristic time of the 
quench $\tau$ we identify three different regimes. For fast quenches the residual energy 
$E_{\mathrm{res}}$ is almost independent on $\tau$. For slower quenches a second intermediate 
region appears in which a power-like decay emerges with $E_{\mathrm{res}} \sim \tau^{-3/2}$. 
By further slowing the quench rate, we find a third large-$\tau$ regime characterized by a different 
power-law, $E_{\mathrm{res}} \sim \tau^{-2}$. All these findings can be accounted for  
by means of an effective Landau-Zener approximation for the finite size LMG model. We complete 
our description of the adiabatic dynamics of the LMG model through the analysis of the 
entanglement entropy of the evolved state.
%}
\end{abstract}

%\pacs{
%}

\maketitle

\section{Introduction}
\label{intro}

Understanding the adiabatic quantum dynamics of many-body systems is central to many 
areas of physics and information science. In adiabatic quantum computation~\cite{Farhi_SCI01}, 
or in quantum annealing (see [\onlinecite{Santoro_JPA:review}] for a review) the ultimate goal 
is to find the ground state of a complex system by adiabatically transforming the underlying 
Hamiltonian. Indeed any quantum algorithm can be efficiently implemented through the 
adiabatic evolution of a system from an initial exactly known state toward a unknown final one 
in which encodes the answer to the specific computational task~\cite{Aharonov:proceeding}.
Once the time scale on which the Hamiltonian is varied is large compared to the typical inverse 
spectral gap of the system, the quantum adiabatic theorem~\cite{Messiah:book} ensures that 
%{\color{red}
if the system was prepared in the ground state of the initial Hamiltonian, at the end of the 
evolution the quantum state 
will be the ground state of the final Hamiltonian.
%} 
The bottleneck to the speed at which the 
algorithm is performed is thus given by those places where the instantaneous Hamiltonian has a 
spectrum where the gap closes in the thermodynamic limit, i.e. as the number of qubits increases. 
If the minimum gap closes faster than a power of the number of qubits then the corresponding 
computational task is intractable. 

The closing of a gap between the ground state and the first excited level in the thermodynamic 
limit is a distinct feature of second order quantum phase transition. It is responsible for the 
critical slowling down~\cite{Sachdev:book}, and the evolution becomes necessarily not adiabatic. 
The problem of adiabatic dynamics close to a critical point, and the consequent defects formation, 
has  originally born in the study of phase transitions in the early universe~\cite{Kibble:review,Zurek:review}. 
The recent extension to the quantum case~\cite{Zurek_PRL05,Polkovnikov_PRB05} has stimulated  
an intense theoretical activity~\cite{Damski_PRL05,Dziarmaga_PRL05,Cherng_PRA06,
Dziarmaga_PRB06,Schutzhold_PRL06,Cincio_PRA07,Cucchietti_PRA07,Fubini_NJP07,
Polkovnikov_07:preprint,Damski_PRL07,Lamacraft_PRL07,Caneva_PRB07,Sengupta_PRL08,Pellegrini_PRB08,
Mukherjee_PRB07,Divakaran_08:preprint,Patane_08:preprint,Dziarmaga_08:preprint,Sen_PRL08,Barankov_08:preprint}.

%{\color{red}
In the search for a deeper understanding of the loss of adiabaticity on crossing 
a quantum critical point an important role is played by exactly solvable models. 
The study of non-equilibrium many-body system is indeed a formidable task
and the help of a tractable exactly solvable system can be of great help in testing 
approximate approaches, besides being of interest in itself.
%}
Most of the work done so far in this direction concentrated 
on one-dimensional quantum systems with short range interaction. In this paper we would like 
address a complementary limit, i.e. a model with infinite coordination (in the thermodynamic limit),
but still amenable of an exact solution: the Lipkin-Meschkov-Glick model (LMG). First introduced
by Lipkin, Meschkov and Glick~\cite{Lipkin_NP65} in the context of nuclear physics, it was then 
adopted by the condensed matter community as paradigm of an infinitely coordinated 
solvable system~\cite{Botet_PRL82}. The result of a sudden quench in this model was recently 
discussed in~\cite{Das_PRB06}, here we present results in the opposite regime in which the system 
is dragged adiabatically through the critical point. As it will be shown in the following, although 
the phase transition is of mean field nature, the dynamics leads to non-trivial results. 

The paper is organized as follows: In Sec.~\ref{model_sec} we introduce the Lipkin-Meschkov-Glick 
model and briefly review its properties which are important for the purposes of this work. In the 
same section we also discuss how we solve numerically the dynamics, Sec.~\ref{num_meth_sec},
and the observables  used to quantify the departure from the adiabatic ground state,
Sec.~\ref{res_en_magn_lack_sec} and Sec.~\ref{ententropy}.
 In this work we use the residual energy (the excess energy as 
compared to the adiabatic limit), the incomplete magnetization (the deficit magnetization as 
compared to the adiabatic limit) and the entanglement entropy. The numerical results together 
with the corresponding scaling arguments are presented  in Sec.~\ref{results_sec}. 
In the final section we present a critical assessment of our findings.

\section{The Model} 
\label{model_sec}

The properties of the LMG model have been thoroughly scrutinized in the literature (see e.g. 
[\onlinecite{Botet_PRB83,Pan_PL99,Links_JPA03,Castanos_PRB06,Unanyan_PRL03,
Dusuel_PRL04,Leyvraz_PRL05,Dusuel_PRB05,Ortiz_NP05,Chen_NJP06,Heiss_JPA06,Ribero_PRL07,
Rosensteel_JPA08}] and references therein). Below we briefly recall few results that are 
relevant to the present paper. 
The LMG Hamiltonian describes a system of spins (1/2 in this work) interacting through 
an infinite-range exchange coupling and immersed in a transverse field. Assuming that the 
field is directed along the z-direction the Hamiltonian can be written as
\begin{eqnarray}
       H=-\frac{2}{N}\sum _{i<j}(S_i ^x S_j ^x +\gamma S_i ^y S_j ^y)-
          \Gamma \sum _{i}^N S_i ^z ,
\end{eqnarray}
where $N$ is the number of the spins in the system, $S_i$ are the Pauli
operators, $\gamma$ is the anisotropy parameter and $\Gamma$ is the transverse
field. By introducing the total spin operator $\mathcal{\vec{S}}=\sum_i \vec{S}_i$, 
the Hamiltonian can be rewritten, apart from a additive constant, as
$H=-\frac{1}{N}[\mathcal{S}_x^{2}+\gamma \mathcal{S}_y^{2}]-\Gamma \mathcal{S}_z$.
The Hamiltonian hence commutes with $\mathcal{S}^2$ and does not couple states having a
different parity  of the number of spins pointing in the magnetic field
direction: $[H,\mathcal{S}^2]=0$ and $[H, \prod _i S_i ^z]=0$. In the isotropic 
case $\gamma=1$ also the z-component $\mathcal{\vec{S}}$ is conserved, $[H,\mathcal{S}_z]=0$.

In the thermodynamical limit the LMG model undergoes a second order quantum phase 
transition at $\Gamma _c =1$ characterized by mean-field critical 
exponents~\cite{Botet_PRB83}. The magnetization in the $x$-direction (or in the 
$xy$-plane, for $\gamma=1$) vanishes when $\Gamma \rightarrow 1^{-}$ as  
\begin{equation}
m = \left\{
          \begin{tabular}{ll}  
           $(1-\Gamma ^2)^{1/2}$ & \hspace{0.5cm}  $\Gamma \le 1 $ \\
                                &     \\
           0                    & \hspace{0.5cm}  $\Gamma > 1 $
          \end{tabular}
    \right.
\end{equation}
for all values of the anisotropy parameter $\gamma$.
For $\Gamma > \Gamma _c$ and for any $\gamma$ the ground state is non degenerate;
while for $\Gamma < \Gamma _c$ it is doubly degenerate in the thermodynamical
limit for any $\gamma\neq 1$, signaling the 
breaking of the $Z_2$ symmetry. The gap vanishes at the transition as
\begin{equation}
       \Delta = [(\Gamma -1)(\Gamma-\gamma)]^{1/2} \quad\mathrm{for}\; \Gamma \geq 1.
\label{gap_crit_exp_eq}
\end{equation}

For any  finite $N$ both the magnetization and the gap are modified (as any other physical 
observable).  The finite size scaling behavior is  available in literature in all the 
relevant regimes (see, e.g., [\onlinecite{Botet_PRB83,Dusuel_PRB05}]). The deviation from the 
thermodynamic limit for the gap $\delta \Delta_N = \Delta_N - \Delta$ and the magnetization 
$\delta m_N = m_N - m$ scale as  
\begin{equation}
\begin{tabular}{lll}  
  $\delta \Delta_N \sim N^{-1}$   & \hspace{0.5cm}$\delta m_N \sim N^{-1/2}$ & \hspace{0.3cm}$\Gamma > 1$ \\
  $\delta \Delta_N \sim N^{-1/3}$ & \hspace{0.5cm}$\delta m_N \sim N^{-1/3}$ & \hspace{0.3cm}$\Gamma = 1$ \\
  $\Delta(N) \sim e^{-aN}$        & \hspace{0.5cm}$\delta m_N \sim N^{-1}$   & \hspace{0.3cm}$\Gamma < 1$
\end{tabular}
\label{scalinggt1}
\end{equation}
for $\gamma>1$ (where $a$ is a constant) and 
\begin{equation}
\begin{tabular}{lll}  
  $\delta \Delta_N \sim N^{-1}$          & \hspace{0.5cm}$\delta m_N \sim N^{-1/2}$  & \hspace{0.3cm}$\Gamma > 1$ \\
  $\delta \Delta_N \sim N^{-1}$       & \hspace{0.5cm}$\delta m_N \sim N^{-1/2}$ & \hspace{0.3cm}$\Gamma = 1$ \\
  $\delta \Delta_N \sim N^{-1}$       & \hspace{0.5cm}$\delta m_N \sim N^{-1}$   & \hspace{0.3cm}$\Gamma < 1$
\end{tabular}
\label{scalingeq1}
\end{equation}
for $\gamma=1$, respectively. The scaling behavior of the gap is important in order to distinguish 
the various dynamical regimes in the adiabatic annealing. It is however important to stress at 
this point that the equilibrium gap is not necessarily the one responsible for the loss of 
adiabaticity. As we will see in the following section, due to the parity conservation the 
relevant gap for the dynamics is different from the equilibrium one (although with the same 
scaling behavior). 

%.....................................................................................
\subsection{Adiabatic dynamics} \label{num_meth_sec}
%.....................................................................................

The adiabatic dynamics is implemented by changing the external transverse field from 
an initial value $\Gamma \gg 1$ at $t_{\rm in}=-\infty$, where the ground state of $H(t_{\rm in})$ 
is completely dominated by the transverse field term with all the spins aligned along the
$+\hat{z}$ direction, to $\Gamma =0$, where the ground state is ordered in the $xy$ plane.
The annealing time is characterized by a time scale $\tau$. More specifically we consider the 
case, as often in this type of problems, to reduce the magnetic field linearly in time 
\begin{equation}
  \Gamma (t)=-t/\tau\quad \mbox{for} \quad t\in (-t_{\rm in} ,0]
\end{equation}
with  $t_{\rm in} \gg \tau$.

The problem we want to discuss is further simplified by the following observation. The initial 
state, the ground state of $H(t_{\rm in})$, belongs to the sector of maximum spin $\mathcal{S}=N/2$. 
Since $\mathcal{S}$ is a constant of motion it is sufficient to restrict our attention to this 
subspace only. From now on we assume $\mathcal{S}=N/2$ (for simplicity we consider $N$ even). 
In the basis  $|N/2,\mathcal{S}^z\rangle$  ($\mathcal{S}^z=-N/2,...,N/2$), the Schr\"odinger evolution of 
the state
\begin{equation}
  |\psi (t)\rangle = \sum _{j=1}^{N/2+1} u_{2j-1} (t) |N/2,-N/2-2+2j\rangle ,
\end{equation}
amounts to solving the following set of coupled equations
\begin{equation} \label{Sch_tdep:eqn}
  \displaystyle i\frac{du_{2j-1}}{dt}=\sum _{k}A_{j,k}u_{2k-1}(t) \;.
\end{equation}
The odd amplitudes $|N/2,-N/2-1+2j\rangle$ do not couple because of parity conservation. 
In Eq.(\ref{Sch_tdep:eqn}) $A$ is a $(N/2+1)\times (N/2+1)$ symmetric matrix whose non-zero entries
are given by
\begin{eqnarray}
  A_{j,j+1}  &=& 
  - \frac{1}{4N}(1-\gamma)a_{-N/2-2+2j}a_{-N/2+2j-1} \nonumber \\
  A_{j,j}    &=& \displaystyle -\frac{1}{4N}(1+\gamma)[a^2_{-N/2-3+2j}+a^2_{-N/2-2+2j}]\nonumber\\
              &&-\Gamma (-\frac{N}{2}-2+2j)+\frac{1}{4}(1+\gamma) \;,
\end{eqnarray}
in terms of the usual angular momentum raising operator matrix elements:
\begin{eqnarray}
  a_j=\left[\frac{N}{2}(\frac{N}{2}+1)-j(j+1)\right]^{1/2} \;\; .
\end{eqnarray}
Special values have the boundary terms of $A$, given by:
\begin{eqnarray}
  A_{1,1}&=&\displaystyle -\frac{1}{4N}(1+\gamma)a^{2}_{-N/2}-\Gamma (-\frac{N}{2})\nonumber\\
             &&+\frac{1}{4}(1+\gamma)\nonumber\\
  A_{N/2+1,N/2+1}&=&\displaystyle -\frac{1}{4N}(1+\gamma)a^{2}_{N/2-1}\nonumber\\
              &&-\Gamma (\frac{N}{2})+\frac{1}{4}(1+\gamma) \;.
\end{eqnarray}
The equations (\ref{Sch_tdep:eqn}) were integrated via standard numerical methods with initial conditions given 
by the amplitudes of the ground state of $H(t=t_{\rm in})$. 

%.....................................................................................
\subsection{Residual energy and incomplete magnetization} \label{res_en_magn_lack_sec}
%.....................................................................................

A natural way of quantifying the degree of adiabaticity of the evolution is to  measure 
the residual energy, defined as
\begin{equation}
  E_{\rm res}=E_{\rm fin}-E_{\rm gs} \;,
\end{equation}
where $E_{\rm gs}$ is the ground state energy of $H(t_{\rm fin})$, and
$E_{\rm fin}=\langle \psi (t_{\rm fin})|H(t_{\rm fin})|\psi (t_{\rm fin})\rangle$
is the average energy of the final time-evolved state $|\psi (t_{\rm fin})\rangle$.
Obviously $E_{\rm fin}$, and hence $E_{\rm res}$, depends on the annealing time $\tau$;
the slower the evolution the closer the final energy to $E_{\rm gs}$, hence the small the residual energy.

An alternative way of quantifying the degree of adiabaticity of the evolution is in
terms of the incomplete magnetization in the final state, defined by
\begin{eqnarray}
m_{\rm inc}=m_{\rm gs}-m(t)
\end{eqnarray}
where $m_{\rm gs}$ is the static magnetization of the ground state of $\Gamma=0$ and
$m_{\rm gs}$ is the average magnetization of the final evolved state. Following 
Botet {\em et al}~\cite{Botet_PRL82}, the magnetization $m$ has been defined through 
%
%{
%\color{red}
\begin{eqnarray}
  m^2=\frac{4}{N^2}\langle \psi|\mathcal{S}_x^2+\delta _{\gamma,1}\mathcal{S}_y^2|\psi\rangle ,
\label{magnetization_eq}
\end{eqnarray}
%}
where the expectation value can be taken either the ground state, for $m_{\rm gs}$, or in the evolved state,
for $m(t)$.
As discussed in Botet {\em et al}~\cite{Botet_PRL82}, the previous definition differs from 
that of the spontaneous magnetization; however, it is more amenable for finite size 
systems and it reduces to the spontaneous magnetization in the thermodynamic limit.

Since we are dealing with a model where the coupling has an infinite range, the incomplete 
magnetization is an appropriate way to characterize the loss of adiabaticity. In this case
a correlation length characterizing the typical distance between defects, along the lines 
followed for short range models, cannot be introduced.  

In the Ising limit, $\gamma=0$, at $\Gamma(t=0)=0$, the residual energy and the
incomplete magnetization are related, as they both depend only on the average value 
$\langle \psi (t=0)|\mathcal{S}_x^2|\psi (t=0)\rangle$: The residual energy per site can be expressed as
\begin{eqnarray}
  \frac{E_{\rm res}}{N}=-\frac{1}{N^2}\langle \psi (t=0)|\mathcal{S}_x^2|\psi (t=0)\rangle +\frac{1}{4} \;;
\end{eqnarray}
the incomplete magnetization is given by
\begin{eqnarray}
m_{\rm inc}=1-\sqrt{\frac{4}{N^2}\langle \psi (t=0)|\mathcal{S}_x^2|\psi (t=0)\rangle } \;.
\end{eqnarray}
%
% Mi pare che va da se' ... non coincidono piu' ...
%
%For $\gamma \ne 0$ both the residual energy and incomplete magnetization may provide 
%interesting information on the adiabatic dynamics of the LMG model. 

%....................................................
\subsection{Entanglement entropy} \label{ententropy}
%....................................................

In addition to the previous observables, it was recently shown that important information of 
the lack of adiabaticity in the system can be acquired by analyzing the entanglement 
entropy $S$~\cite{Cherng_PRA06,Cincio_PRA07}. The study of entropy and other measures of 
entanglement has been recently intensively studied to characterize both equilibrium and 
non-equilibrium quantum many-body systems (see [\onlinecite{Amico_RMP08}] for a review).
In the case of the LMG model the ground state entanglement entropy was studied 
in~\cite{stockton03,Latorre_PRA05,barthel06,dus,unanyan05}. In the present work we study 
the time evolution of $S$ during an adiabatic evolution.

Given a bipartition of the system in $L$ and $N-L$ spins the entanglement entropy  associated 
to the reduced density matrix of one of the subsystems, say $\rho_L$, is defined as 
\begin{equation}
  S_{L}=-\mathrm{Tr}(\rho _L \log _2\rho _L) \;.
\label{entropy}
\end{equation} 
The entropy $S_L$ measures the entanglement between the $L$ spins and the rest of the system.  

The entanglement entropy is straightforwardly evaluated by noticing that, being the states 
$|\mathcal{S}=N/2,\mathcal{S}_z\rangle$ symmetric under any permutations of the sites 
and being the maximum value of the total spin achievable only with the maximum value of the spin
in each subsystem, the following decomposition holds~\cite{Latorre_PRA05}:
\begin{eqnarray}
|N/2,\mathcal{S}^z\rangle &=& \sum_ {l=0} ^L p_l ^{1/2}|L/2,l-L/2\rangle\nonumber\\
&\otimes & |(N-L)/2,n-l-(N-L)/2\rangle\nonumber \;.
\end{eqnarray}
In the previous decomposition $n$ and $l$ indicate, respectively, the number of up-spins in the system 
and in the partition which defines the $L$ sites. 
The coefficients appearing are defined as $p_l = L! (N-L)!n!(N-n)! / (l! (L-l)! (n-l)! (N-L-n+l)! N!)$.
With the knowledge of the representation of the evolved state in the basis $|N/2,\mathcal{S}^z\rangle$ 
and by the using previous decomposition, it is immediate to trace out the $N-L$ spins to obtain the reduced 
density matrix $\rho _L$, and calculate its entropy. 

%.....................................
\section{Results} \label{results_sec}
%.....................................

The results presented below were obtained by integrating numerically Eq.~(\ref{Sch_tdep:eqn}). 
We verified that, as for the initial time of the evolution, it is enough to consider 
$t_{\rm in}=-5\tau$ for faster sweeps ($1 <\tau < 500$) and $t_{\rm in}=-2\tau$
for slower ones. We checked (data not reported) that our results do not depend on the precise 
value of $t_{\rm in}$.
We considered systems up to $N=1024$ spins and annealing times up to $\tau \sim 10^3 - 10^4$.
%
%---------------------------------------------------------------------------------------------------
\begin{figure}
\epsfig{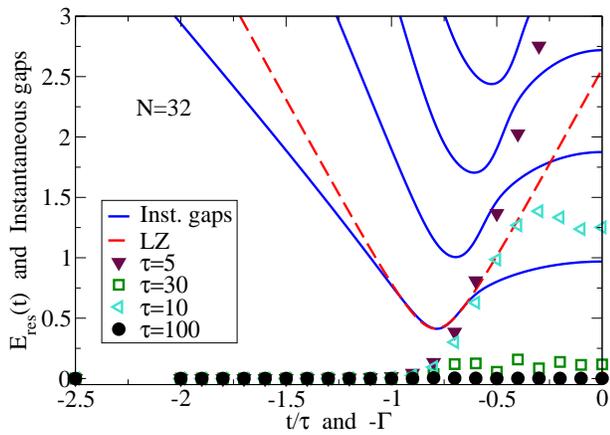}
\caption{(Color online) Residual energy $E_{\rm res}(t)$ versus $t$ for a given instance with 
$N=32,\gamma =0$ of the
LMG model at different values of $\tau$. The solid lines
are the lowest-lying instantaneous spectral gaps as a function of $\Gamma$.
The red-dashed
line is the best fit to the lowest gap used to calculate the Landau-Zener transition rates. 
}
\label{instant_lipkinL32_fig}
\end{figure}
%---------------------------------------------------------------------------------------------------

\begin{figure}
\epsfig{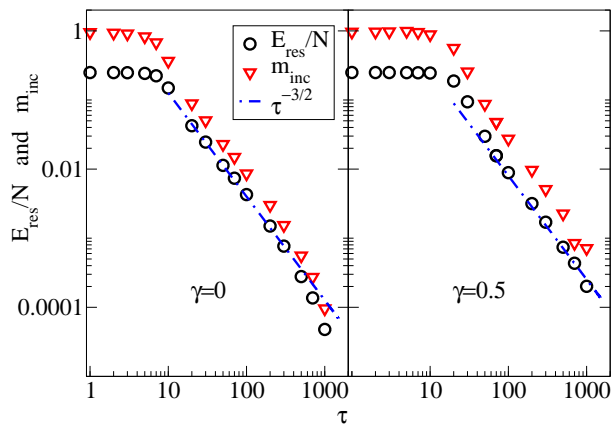}
\caption{(Color online) 
Residual energy per site and incomplete magnetization for the LMG model with $N=1024$ for different values
of the anisotropy parameter $\gamma$. In all cases, for slow enough quenches, a power-law behavior $\tau^{-3/2}$
appears.}
\label{eres_mlac_gam_var_fig}
\end{figure}

\begin{figure}
\epsfig{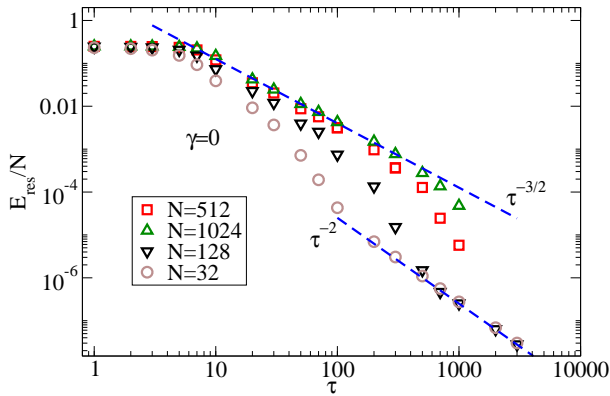}
\caption{(Color online) Residual energy per spin as function of $\tau$
for $\gamma =0$ compared with different power-law behaviors. 
}
\label{E_res_vs_tau_lipkin_Nlarge_fig}
\end{figure}

{\em Residual energy and incomplete magnetization -}
In order to understand the mechanism that leads to breakdown of adiabaticity in the LMG model 
it is instructive to start with one particular example. 
In Fig.\ref{instant_lipkinL32_fig} we chose a system with $N=32$ spins and $\gamma =0$, 
showing the time evolution of the residual energy for different values of the annealing 
time $\tau$. We also plot the instantaneous {\em accessible} gaps (thick solid lines)
obtained by diagonalizing the Hamiltonian at any given $\Gamma$. As one can see, as soon as 
the system loses the adiabaticity, for fast annealing, it starts to ramp up in energy. 
The characteristic time scale for breaking of adiabaticity is however not given by the equilibrium smallest gap.
As noticed in the previous Section, the dynamics is restricted to the subspace with fixed total spin 
$\mathcal{S}=N/2$ and can involve only states with the same parity of $\mathcal{S}^z$~\cite{footnote_XXevolution}.
%
%From the picture emerges that we can locate the minimum and most important gap at the critical point 
%$\Gamma _c =1$, with corrections originating from the finiteness of the system.
%
%---------------------------------------------------------------------------------------------------
\begin{figure}
\epsfig{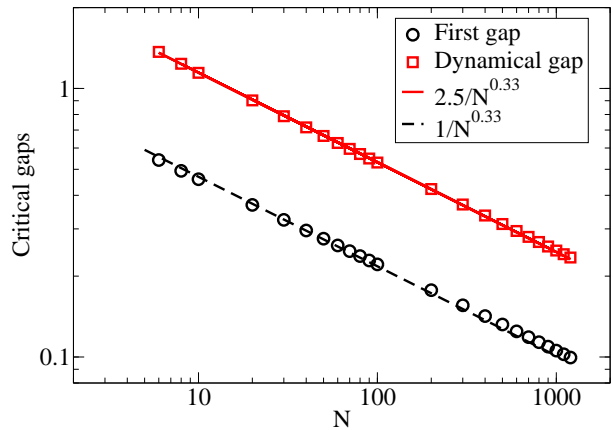}
\caption{(Color online) Smallest gap and dynamical gap at the critical point as function of the size of the 
system.
}
\label{lip_crit_gaps_fig}
\end{figure}
%------------------------------------
%
Hence, the first gap relevant for the dynamics, that we call {\em dynamical} gap, is the energy difference 
between the ground state and the {\em second} excited state, the smallest gap being forbidden by 
parity conservation of the total spin along the $z$-axis.
As shown in Fig.\ref{lip_crit_gaps_fig}, the dynamical gap exhibits the same critical 
behavior of the excitation gap~\cite{Botet_PRB83}: both close polynomially in the thermodynamical 
limit with the same dynamical exponent $z = 1/3$, $\Delta_c\sim N^{-z}$. 
This is usually accompanied by a polynomial-like decay of the residual energy with 
increasing annealing time $\tau$. 
This is indeed the case, for both the residual energy and the incomplete magnetization, 
as shown in Fig.\ref{eres_mlac_gam_var_fig} and, more in detail for $\gamma=0$, in 
Fig.\ref{E_res_vs_tau_lipkin_Nlarge_fig}. 
%
%----------------------------------------------------------------------
%----------------------------------------------------------------------
%
%----------------------------------------------------------------------
%----------------------------------------------------------------------
%
The behavior appears to be qualitatively independent on the value of the anisotropy parameter $\gamma$ for 
$\gamma <1$, see Fig.\ref{eres_mlac_gam_var_fig}; this was expected due to the fact that the minimum gap
has the same large-N behavior irrespective of the anisotropy.
In the following only the case $\gamma =0$ will be discussed.

Inspection of Fig.\ref{E_res_vs_tau_lipkin_Nlarge_fig} reveals three different regimes. 
For fast quenches the dynamics involves almost all the levels, see, e.g., Fig.\ref{instant_lipkinL32_fig} 
for $\tau =5$. The residual energy per site is near to its maximum and
shows very little dependence on the size of the system 
%{\color{red}
and on the annealing time $\tau$.
%} 
For larger values of $\tau$, a second intermediate region appears in which a power-like decay emerges, 
with $E_{\mathrm{res}} \sim \tau^{-3/2}$. 
Finally, by further slowing the quench rate, a third large-$\tau$ regime characterized by 
a different power-law, $E_{\mathrm{res}} \sim \tau^{-2}$, emerges. We briefly discuss the emergence of the 
last two regimes by means of a Landau-Zener approach adapted to the present problem.

The argument follows closely the one given in~\cite{Zurek_PRL05}.
The probability of exciting the system into the first excited state, obtained by the 
Landau-Zener formula 
\begin{equation} \label{lz_formula_eq}
  P_{LZ}\simeq e^{-\alpha \Delta ^2\tau} \;,
\end{equation}
with $\alpha=\pi/4$, gives a lower bound to the true transition probability, 
as it ignore the transitions to all the other excited levels. 
Using the scaling of the critical point gap with the number of spins, $\Delta\sim N^{-1/3}$, 
it is possible to determine maximum system size for a defect-free quench once the probability for this 
to occur is fixed to the value $\tilde{P}_{ex}$.
This gives:
\begin{eqnarray}
  \frac{1}{N_{\rm free}} \sim \left(\frac{|\ln \tilde{P}_{ex}|}{\alpha}\right)^{3/2}\frac{1}{\tau ^{3/2}} \;.
\end{eqnarray}
One can consider $1/N_{\rm free}$ as an estimate of the fraction of the flipped spins after the quench. 
The residual energy per site in the LMG model can then be evaluated to be 
\begin{eqnarray} \label{lz_estimate_eq}
  \frac{E_{\rm res}}{N}\sim \frac{1}{N^2} \frac{N}{N_{\rm free}}N\sim \frac{{\rm const.}}{\tau ^{3/2}} \;.
\end{eqnarray}
This simple estimate is in good agreement with the numerical data in the intermediate regime of 
Fig.\ref{E_res_vs_tau_lipkin_Nlarge_fig}. 

%{\color{red}
For short range models the same power law  of Eq.(\ref{lz_estimate_eq}) can be also derived by determining 
the spatial scale over which defects occur~\cite{Zurek_PRL05}. We tried to apply the arguments of 
Zurek {\em et al}\cite{Zurek_PRL05} to the LMG model by identifying the correlation length with the 
coherence number introduced in~[\onlinecite{Botet_PRL82}]. The procedure we followed, however, does not 
lead to the correct exponent. We have reasonable confidence that the failure in obtaining the 
correct scaling with this second method may be related to the above identification and the consequent 
definition of defect density. It would be interesting to find the correct argument in order to extend 
the approach by Zurek {\em et al}\cite{Zurek_PRL05} or Polkovnikov~\cite{Polkovnikov_PRB05} to infinite 
range models.
%}

\begin{figure}
\epsfig{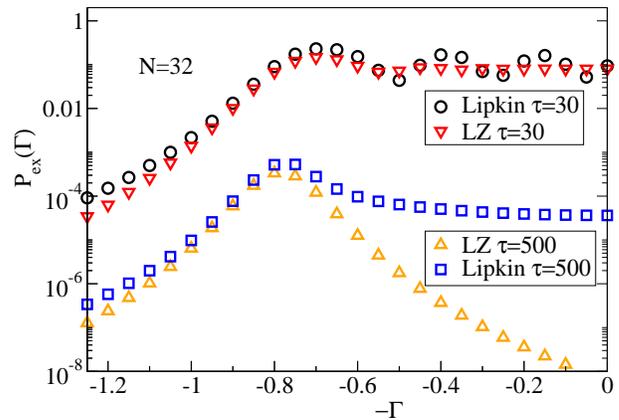}
\caption{(Color online)
Comparison between the excitation probabilities as function of $\Gamma$ of the LMG model with $N=32$ 
and of its effective LZ approximation for different values of $\tau$.}
\label{P_gamma_L32_fig}
\end{figure}

\begin{figure}
\epsfig{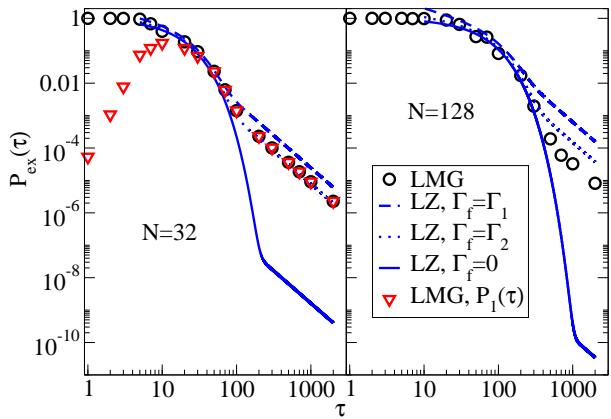}
\caption{(Color online)
Excitation probability as function of $\tau$ for LMG systems (circles) of two different sizes compared to that
one of the respective  
LZ-effective models (line) for different final times. For the LZ models the first two terms of the Vitanov 
approximation have been used. For the case LMG N=32, the probability of exciting the first level is 
also presented (triangles).}
\label{P_tau_bound_fig}
\end{figure}

{\em Effective two-level approximation -}
As already mentioned before there is, for slower quenches, a further crossover to a different power-law. 
Can one explain also this behavior by using a Landau-Zener argument? To this end, it is important 
to refine this comparison and to understand to which extent the dynamics of a many-body system 
described by the LMG model can be described by two (many-body) levels. 
In general, in a many-body system there will be a number of avoided crossings and multiple LZ transitions, 
including interference between them.
Only when a single avoided crossing is dominant and well separated from the others a two-level 
approximation is appropriate. 
A detailed analysis of this issue is summarized in Figs.\ref{P_gamma_L32_fig} and~\ref{P_tau_bound_fig}
where we show the case of $N=32$ as an example. Our analysis starts by extracting 
the best dynamical minimum gap and adapting to it the following two-level Hamiltonian:
\begin{eqnarray}
H_{LZ}=\left(
\begin{array}{cc}
-\Omega _{LZ}( \Gamma-\Gamma _0) & \Delta _{LZ} \\
\Delta _{LZ}     &  \Omega _{LZ}(\Gamma -\Gamma _0) \;.
\end{array} 
\right),
\label{lz_ham_eq}
\end{eqnarray}
%
%{\color{red}
In the effective Landau-Zener problem $\Omega _{LZ}, \Delta _{LZ}$ and $\Gamma _0$ are the fitting 
parameters and $\Gamma =-t/\tau$.
%} 
In Fig.\ref{instant_lipkinL32_fig} the dashed line represents the instantaneous gap 
of the Hamiltonian (\ref{lz_ham_eq}) suited to the case $N=32$. 
From here we compare the results of the full LMG model with those obtained using 
%the fitted gap and 
LZ theory. 
%
%----------------------------------------------------------------------
%----------------------------------------------------------------------
%
As shown in Fig.\ref{P_gamma_L32_fig}, the excitation probability in the LMG model 
for slow enough quenches coincides with that of the effective LZ problem. 
It appears that this approximation is good also in the estimate of the asymptotic value of the probability 
for $ 10<\tau <100$. Deviations come predominantly from the more enhanced oscillations of the 
post crossing region in the LMG model.
%{\color{red}
For larges $\tau$'s the asymptotic value obtained from the effective two-level system gives a very poor 
approximation to the actual data. This can be traced back to the presence of further crossings which 
are obviously neglected in the two-level approximation. In our LZ scheme this can be effectively corrected 
by approximating the LZ crossing probability to the time before the next level crossing comes into play. 
This is explained below
%} 
 
As found by Vitanov~\cite{Vitanov_PRA99}, it is possible to define the duration of a single LZ as the 
time required by the probability for jumping from zero to its asymptotic value, linearly and with the 
slope calculated at the crossing point.
Using $\Gamma$ as time-scale one can write:
\begin{equation}
  \Gamma _{\mathrm{jump}}\sim \frac{P(\infty)}{P'(\Gamma _{\mathrm{cross}})} \;.
\end{equation}
This time turns out to be \emph{exponentially} divergent with $\tau$ for large $\tau$~\cite{Vitanov_PRA99}.
This means that for slow quenches consecutive LZ transitions are not independent. 
In a first crude approximation, we can guess that the consequence of this is simply to stop the 
probability from relaxing towards the asymptotic value when the system has reached the second crossing.
%
%---------------------------------------------------------------------------------------------------
%---------------------------------------------------------------------------------------------------
%
The presence of a power-law regime $\sim \tau^{-2}$ for extremely slow dynamics is 
a clear consequence of the finite duration of the evolution. 
In the original works by Landau and by Zener, the final time is supposed to be $t_f=\infty$; 
here the evolution is stopped at $\Gamma_f=-t_f/\tau =0$ for the LMG model, and at 
$t_{(LZ)f}=-\Gamma_{(LZ)f}\cdot\tau$ for the effective LZ, with $\Gamma _{(LZ)f}=\Gamma _f -\Gamma _0$. 
%
%by assuming that the avoided level crossing take place at $t_{(LZ)}=0$, 
%see Fig.(\ref{instant_lipkinL32_fig}) -- the two 
%intrinsic time are shifted each other of the distance of minimum gap from zero in the LMG model.\\
%
An accurate analysis of the finite-time Landau-Zener model (FTLZ) has been done in 
Ref.~[\onlinecite{Vitanov_PRA99}], where it is shown that the transition probability reads, 
in this case~\cite{footnote_Vitanov}: 
\begin{eqnarray}
P_{(FTLZ)}(\tau) &\sim & P_{LZ}(\tau)\nonumber\\
%&+&\frac{(1-2P_{LZ}(\tau))}{16\tau ^2 \Delta _{LZ} ^4(1+(\Gamma _{(LZ)f}/\Delta _{LZ})^2)^3}\nonumber\\
&+&\frac{(1 - 2P_{LZ}(\tau))}{16\Delta _{LZ} ^4\frac{\tau ^2}{\Omega _{LZ} ^2}(1+\frac{\Omega _{LZ} ^2}{\Delta _{LZ} ^2}\Gamma _{(LZ)f} ^2)^3}
\label{vitanovLZ}
\end{eqnarray}
with $P_{LZ}(\tau)=e^{-\pi\Delta _{LZ} ^2\tau/\Omega _{LZ}}$.
%, $\Delta_{LZ}$ being half of the gap at 
%the anticrossing point and $\Gamma_{(LZ)f}$ representing the final time in unit of $\tau$.
 As it can be immediately seen from the previous 
equation, by sending the final time to infinity the usual LZ probability is recovered.
%
%{\color{red} 
The crossover rate $\hat{\tau}$ to the $\tau^{-2}$ scaling is obtained by equating 
the two terms on the r.h.s. of Eq.(\ref{vitanovLZ}).  In the limit 
%$16\frac{\Delta _{LZ} ^4}{\Omega _{LZ} ^2}(1+\frac{\Omega _{LZ} ^2}{\Delta _{LZ} ^2}\Gamma _{(LZ)f} ^2)^3 \ll 1$
$\frac{8}{\pi}(1+\frac{\Omega _{LZ} ^2}{\Delta _{LZ} ^2}\Gamma _{(LZ)f} ^2)^{3/2}\gg 1$
the crossover time is approximated by 
\begin{equation}
\hat{\tau} \sim 
%\frac{2}{\pi} \frac{\Omega _{LZ}}{\Delta _{LZ} ^2} 
%\ln \left[\frac{1}{\frac{8}{\pi}(1+\frac{\Omega _{LZ} ^2}{\Delta _{LZ} ^2}\Gamma _{(LZ)f} ^2)^{3/2}}\right]
\frac{\Omega _{LZ}}{4\Delta _{LZ}^2}\frac{1}{(1+\frac{\Omega _{LZ} ^2}{\Delta _{LZ} ^2}\Gamma _{(LZ)f} ^2)^{3/2}}.
\label{crosstau}
\end{equation}
%
%}
In Fig.\ref{P_tau_bound_fig} 
we compared the excitation probabilities of LMG systems of different sizes with their 
single-LZ approximations. The probabilities for the effective models are evaluated for three different
final time: $\Gamma_f=0,\Gamma_{1},\Gamma_{2}$, where the last two are the positions, respectively, 
of the minimum gap between the ground state and the second excited level, and the minimum gap between 
the first and the second excited levels. As it can be seen, the agreement is quite good and one can 
reproduce in this way also the regime with the $\tau^{-2}$ behavior.

\begin{figure}[t]
\epsfig{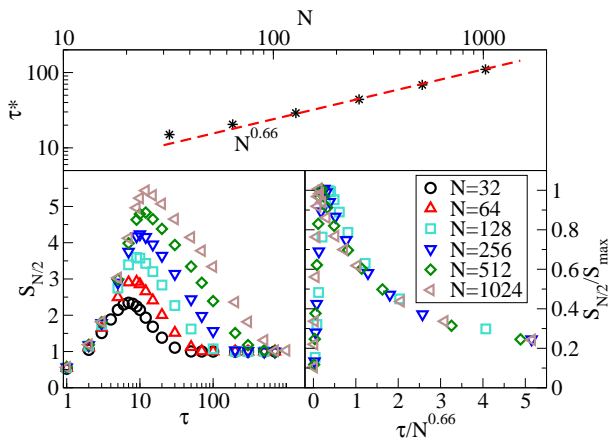}
\caption{(Color online) Left lower panel: entanglement entropy of a block of $L=N/2$ spins as function 
of the quench time $\tau$. Right lower panel: entanglement entropy of a block of $L=N/2$ divided by 
its maximum value as function of the rescaled variable $\tau /N^{0.66}$. Upper panel: the time 
scale $\tau ^*$, see the text for the definition, as function of system size $N$.
}
\label{ent_vs_tau_fig}
\end{figure}

{\em Entanglement entropy -}
We finally would like to discuss the behavior of the entanglement entropy which was already 
used as a tool to characterize adiabatic many-body dynamics in Refs.~[\onlinecite{Cherng_PRA06,Cincio_PRA07}].
Our results are summarized in Fig.\ref{ent_vs_tau_fig}. In the left lower panel  
the entanglement entropy, for a block of size $L=N/2$, of the state evolved down to $\Gamma_f=0$ 
is plotted as function of the quench time $\tau$. 
%In all the curves the block of spins over which the entropy has been calculated is $L=N/2$. 
For fast quenches, $\tau\to 0$,the state does not evolve (it remains in a nearly factorized state), 
thus the entanglement necessarily tends towards zero.
For very slow dynamics $\tau\to \infty$, since we are dealing with finite systems, the evolution 
eventually becomes adiabatic and the entanglement picks up the value it assumes in the final ground state, 
$S_{\mathrm{gs}}(\Gamma_f=0)=1$, independently on the subsystem size~\cite{Latorre_PRA05}.
Between this two limiting behaviors, the entropy reaches a size-dependent maximum at an intermediate
value of $\tau$.
%
%---------------------------------------------------------------------------------------------------
%---------------------------------------------------------------------------------------------------
%
An interesting feature is that the presence of a finite minimum gap can be easily 
connected with a time scale for the decaying of the entanglement.
A possible choice for this time scale consists in selecting the $\tau^*$ at which the 
entropy has reduced by half the value of the its peak respect to the slow quench limit
\begin{eqnarray}
  S_{N/2}(\tau^*) = \frac{(S_{\mathrm{max}}-1)}{2}+1 \;.
\end{eqnarray}
In the upper panel of Fig.~(\ref{ent_vs_tau_fig}), $\tau^*$ determined in this way is shown as 
a function of the system size $N$. 
For large $N$, a power-like behavior emerges with an exponent $\sim 0.66$, hinting at a relation
\begin{equation} \label{lz_gap_relation_eq}
  \tau^* \sim \frac{1}{\Delta ^2} \;.
\end{equation}
In the lower right panel of Fig.~(\ref{ent_vs_tau_fig}), the entanglement $S_{N/2}$ divided 
by its maximum value $S_{\rm max}$ is plotted as a function of the rescaled variable $\tau/N^{0.66}$, 
showing, for large systems ($N\geq 128$), a collapse of all data on the same curve.
Note that Eq.~(\ref{lz_gap_relation_eq}) expresses exactly the same energy-time relation 
found in the usual LZ system, see Eq.(\ref{lz_formula_eq}), 
so that the correspondence stated in previous sections is again supported. % Che vuol dire? 

%........................................................
\section{Conclusions} \label{conclusions_sec}
%........................................................

In this paper we have studied the adiabatic quantum dynamics of the LMG model in a transverse field 
across its quantum critical point. 
We focused our attention on the residual energy after the quench analyzing its behavior 
as a function of the annealing time, in order to evaluate the extent of diabaticity of the evolution.
The dynamics is restricted to a subspace of definite total spin and parity of its 
projection along the $z$-axis, due to the symmetries of the Hamiltonian. 
Results appeared to be qualitatively independent of the value of the $XY$-anisotropy parameter 
$\gamma$, except for the fully isotropic $XX$ case at $\gamma =1$, where the further conservation of 
$\mathcal{S}^z$ plays an important role.
By starting the evolution in the ground state for very large values of the transverse field $\Gamma$, 
and then reducing $\Gamma(t)$ linearly to zero, three regimes in the residual 
energy are identifiable: the first one, corresponding to fast quenches, is strongly not adiabatic,
involves transitions from the ground state towards many excited states and is characterized by 
a residual energy near to its saturation value. 
In the intermediate regime, the lowest critical dynamically accessible gap starts dominating the evolution, 
inducing a residual energy per site that decays in a power-like manner, like $\tau^{-3/2}$.
%{\color{red}
The third large-$\tau$ region, where the residual energy decays like $\tau^{-2}$, is understood 
by taking into account the presence of additional level crossings. In the effective Landau-Zener 
description we used in this paper this results in the requirement to consider a  
finite-time Landau-Zener sweep. As show by Vitanov a finite-time sweep leads to a polynomial (in $\tau$)
contribution to the LZ transition probability which is dominant for very slow sweeping rates.  
Notice that this $\tau^{-2}$ regime, usually described as the general deviation from adiabaticity 
deriving by the adiabatic theorem for very slow evolutions \cite{footnote_Suzuki}, 
emerges here in an alternative way through the parallelism with an effective FTLZ model.
%}

\acknowledgments
This work has been supported by MIUR-PRIN and EC-Eurosqip. We acknowledge fruitful 
discussion with Luigi Amico, Elena Canovi, Simone Montangero, Dario Patan\'e, and 
Alessandro Silva.

%%%%%%%%%%%%%%%%%%%%%%%%%%%%%%%%%%%%%%%%%%%%%%%%%%%%%%%%%%%%%%%%%%%%%%%%%
%                               BIBLIOGRAPHY
%%%%%%%%%%%%%%%%%%%%%%%%%%%%%%%%%%%%%%%%%%%%%%%%%%%%%%%%%%%%%%%%%%%%%%%%%
\bibliographystyle{apsrev}
%\bibliography{QA}

\begin{thebibliography}{56}
\bibitem[{foo({\natexlab{a}})}]{footnote_XXevolution}
\bibinfo{note}{The case $\gamma=1$, due to the conservation of $J^z$, presents
  only a trivial evolution for that concerns our analysis and so has not been
  considered.}
\bibitem[{foo({\natexlab{b}})}]{footnote_Vitanov}
\bibinfo{note}{There is also a third term in the formula of Vitanov, but it is
  negligible respect to the second one in the large $\tau$ limit, for details
  refer to Ref[\onlinecite{Vitanov_PRA99}].}
\end{thebibliography}

\end{document}